\documentclass[a4paper]{article}
\usepackage{INTERSPEECH2022, caption, multirow, makecell, array, color, colortbl}
\definecolor{Gray}{gray}{0.9}
\title{Asymmetric Proxy Loss for Multi-View Acoustic Word Embeddings}
\name{Myunghun Jung, Hoirin Kim}
\address{School of Electrical Engineering, KAIST, Daejeon, Republic of Korea}
\email{kss2517@kaist.ac.kr, hoirkim@kaist.ac.kr}
\begin{document}
\maketitle
%
\begin{abstract}
Acoustic word embeddings (AWEs) are discriminative representations of speech segments, and learned embedding space reflects the phonetic similarity between words.
With multi-view learning, where text labels are considered as supplementary input, AWEs are jointly trained with acoustically grounded word embeddings (AGWEs).
In this paper, we expand the multi-view approach into a proxy-based framework for deep metric learning by equating AGWEs with proxies.
A simple modification in computing the similarity matrix allows the general pair weighting to formulate the data-to-proxy relationship.
Under the systematized framework, we propose an asymmetric-proxy loss that combines different parts of loss functions asymmetrically while keeping their merits.
It follows the assumptions that the optimal function for anchor-positive pairs may differ from one for anchor-negative pairs, and a proxy may have a different impact when it substitutes for different positions in the triplet.
We present comparative experiments with various proxy-based losses including our asymmetric-proxy loss, and evaluate AWEs and AGWEs for word discrimination tasks on WSJ corpus.
The results demonstrate the effectiveness of the proposed method.
\end{abstract}
%
\section{Introduction}
\label{sec:1}
Deep metric learning is an approach to nudge a deep neural network to represent the desired embedding space where similar samples are guided to be embedded closer and dissimilar ones to be pushed away from each other.
Depending on the criterion to judge whether given pair is similar or dissimilar, deep metric learning can be categorized into various tasks and have a different meaning of embeddings accordingly.

Here, unlike the terminology learning contextual semantics for natural language processing \cite{mikolov2013distributed}, word embeddings that reflect the phonetic similarity between words are called \textit{acoustic word embeddings} (AWEs).
With its discriminative property, AWEs are mainly employed to word discrimination tasks such as query-by-example spoken term detection \cite{levin2013fixed, chung2016audio, kamper2016deep, settle2016discriminative} and open-vocabulary wake-up word detection \cite{lim2019interlayer}.

For learning AWEs, there have been several works that explored Siamese networks based on Triplet loss.
In \cite{kamper2016deep, bengio2014word, yuan2018learning}, authors used convolutional neural networks (CNNs) to transform variable-length of speech segments into fixed-dimensional embeddings.
Soon, a benchmark study \cite{settle2016discriminative} found that types of recurrent neural networks (RNNs), such as long short-term memory (LSTM) and gated recurrent unit (GRU), are more suitable as they utilize a mechanism for selectively retaining or discarding temporal features \cite{chen2015query, settle2017query}.
A hierarchical structure of LSTMs connected by an attention layer \cite{lim2019interlayer} was also introduced to consider local and global contexts.

These single-view approaches, meaning that only speech segments are considered as input, are inevitably constrained in available amounts of information.
To overcome this problem, a pioneering study \cite{he2017multi} adopted multi-view learning where the networks are trained with two types of data, speech segment and its corresponding text label.
As well as the network for AWEs, another network is jointly trained to transform text labels into \textit{acoustically grounded word embeddings} (AGWEs) \cite{settle2019acoustically, hu2020multilingual}.
The fact that the orthography of text labels naturally reflects the phonetic similarity between their pronunciations allows AWEs and AGWEs to be projected onto the same embedding space.
By drawing complex information from supplementary data, the multi-view approaches have shown superior performance.

In our prior work \cite{jung2019additional}, we analyzed the grounds for improvement of the multi-view approaches by focusing on the uniqueness of AGWEs.
Given one word, its verbalized instances appear differently so that the network produces slightly different AWEs, which can bring about potential intra-class variance.
In contrast, its written word is always unique, resulting in only one AGWE for each word class.
Then the AGWE plays a key role in training as a pivot point for the positive AWEs to be easily centralized.
From these characteristics, we found some analogies between AGWEs and \textit{proxies} in deep metric learning.

Introduced in \cite{movshovitz2017no}, a proxy is an element of a concise set that approximates all data points.
As a representative of each data class, a proxy can replace the relationship between data-to-data as one between data-to-proxy, which leads to a significant reduction in training complexity.
For example, pair-based losses such as NCA \cite{goldberger2004neighbourhood}, N-pair \cite{sohn2016improved}, and Multi-Similarity (MS) \cite{wang2019multi} losses were reformulated into proxy-based losses respectively in \cite{movshovitz2017no, aziere2019ensemble, kim2020proxy} by simply modifying the ways to construct a batch and to compute a similarity matrix.

In this paper, we expand the multi-view approach into a proxy-based framework for deep metric learning by equating AGWEs with proxies.
Based on the general pair weighting (GPW) \cite{wang2019multi}, we systematize the proxy-based framework according to the composing functions, the position of proxies in a triplet, and whether to apply the prior two factors in asymmetry.
Most works following the GPW have used the same formula for both anchor-positive and anchor-negative terms.
On the contrary, we argue that the optimal function making anchor and positive closer may differ from one for anchor and negative to be farther.
Also, we assume that proxies have different impacts when they substitute for anchor, positives, or negatives in a triplet.
So, we propose an Asymmetric-Proxy loss that combines different parts of loss functions asymmetrically while keeping their merits and considers the position of proxies differently in computing the similarity matrix.

We evaluate learned AWEs and AGWEs on Wall Street Journal corpus for word discrimination tasks.
Under the proxy-based framework, we present comparative experiments with various deep metric learning losses, including proposed Asymmetric-Proxy loss and its possible candidates.
The results demonstrate the effectiveness of the proposed method.
%
\section{Pair-based framework}
\label{sec:2}
\subsection{Problem setting}
\label{ssec:2.1}
Let $\left\{ \left( x_i, c_i \right) | 1 \leq i \leq N \right\}$ be a batch of $N$ data, where $x_i$ is a speech segment and $c_i$ is a word class.
An acoustic encoder $f(\cdot; \theta)$ projects $x_i$ onto a $d$-dimensional embedding space, resulting in AWE $f(x_i; \theta)$.
For each data index $i$, batch indices divide into the set of positives $\mathcal{P}_i = \left\{ j | c_j = c_i, 1 \leq j \leq N \right\}$ and the set of negatives $\mathcal{N}_i = \left\{ j | c_j \neq c_i, 1 \leq j \leq N \right\}$.
We use the cosine similarity matrix $\mathbf{S} \in \mathbb{R}^{N \times N}$ of all embedding pairs.
Then a pair-based framework has the following objective for all $1 \leq i \leq N, j \in \mathcal{P}_i,$ and $k \in \mathcal{N}_i$:
\begin{equation}
S_{ij} > S_{ik}.
\label{eq:1}
\end{equation}
\subsection{General pair weighting}
\label{ssec:2.2}
The general pair weighting (GPW) \cite{wang2019multi} is a generalized pair-based framework, where various pair-based losses to optimize the objective in Eq.~(\ref{eq:1}) can be formulated as a summation of anchor-positive term $\mathcal{F}(\mathbf{S}, \mathcal{P})$ and anchor-negative term $\mathcal{F}(\mathbf{S}, \mathcal{N})$ as follows:
\begin{equation}
\mathcal{L} = \sum_{i = 1}^{N} \Big( \mathcal{F}(\mathbf{S}, {\mathcal{P}_i}) + \mathcal{F}(\mathbf{S}, {\mathcal{N}_i}) \Big).
\label{eq:2}
\end{equation}
Here, $\mathcal{F}$ is a function of similarities and their gradients have to satisfy the following condition:
\begin{equation}
\frac{\partial \mathcal{F}(\mathbf{S}, \mathcal{P})}{\partial S_{ij}} \leq 0 \quad \mathrm{and} \quad \frac{\partial \mathcal{F}(\mathbf{S}, \mathcal{N})}{\partial S_{ik}} \geq 0.
\label{eq:3}
\end{equation}

Unfortunately, there are so many functions that meet the condition.
Among them, we introduce two functions that have been widely used in recent works.
So rather than handling all functions in generalities, we compromise our research scope within them.
\subsection{Mean softplus function}
\label{ssec:2.3}
The first one is \textit{Mean-SoftPlus} (MSP) function.
Softplus \cite{dugas2000incorporating} is a kind of activation function that optimizes the binomial log-likelihood \cite{hastie2009elements}, and is adopted in Binomial Deviance (BD) loss \cite{yi2014deep} for deep metric learning.
When we organize the loss like in Eq.~\ref{eq:2}, it is formulated as summation of two MSP functions:
\begin{equation}
\begin{split}
\mathcal{L} &= \sum_{i = 1}^{N} \left( \frac{1}{|\mathcal{P}_i|} \sum\limits_{j \in \mathcal{P}_i} \mathrm{log} \left( 1 + e^{\alpha (\lambda - S_{ij})} \right) \right.\\
& \qquad \qquad + \left. \frac{1}{|\mathcal{N}_i|} \sum\limits_{k \in \mathcal{N}_i} \mathrm{log} \left( 1 + e^{\beta (S_{ik} - \lambda)} \right) \right)\\
&= \sum_{i = 1}^{N} \left( \underset{j \in \mathcal{P}_i}{\mathrm{MSP}} \, \alpha(\lambda - S_{ij}) + \underset{k \in \mathcal{N}_i}{\mathrm{MSP}} \, \beta(S_{ik} - \lambda) \right),
\end{split}
\label{eq:4}
\end{equation}
where $\alpha$ and $\beta$ are scaling hyper-parameters, and $\lambda$ is a margin.

Due to the linear combination of softplus functions, MSP expresses the \textit{self-hardness} shown in the following gradients:
\begin{equation}
\left\{ \begin{aligned}
\frac{\partial \mathcal{L}}{\partial S_{ij}} = \frac{\alpha}{|\mathcal{P}_i|} & \frac{- e^{- \alpha S_{ij}}}{e^{- \alpha \lambda} + e^{- \alpha S_{ij}}}, & j \in \mathcal{P}_i\\
\frac{\partial \mathcal{L}}{\partial S_{ik}} = \frac{\beta}{|\mathcal{N}_i|} & \frac{e^{\beta S_{ik}}}{e^{\beta \lambda} + e^{\beta S_{ik}}}, & k \in \mathcal{N}_i
\end{aligned} \right. .
\label{eq:5}
\end{equation}
From the sigmoidal form of gradients, the higher weight is assigned to the more violating pair, where the similarity itself determines the magnitude in Eq.~(\ref{eq:5}).
\subsection{Extended LogSumExp function}
\label{ssec:2.4}
The next is \textit{Extended-LogSumExp} (ELSE) function.
ELSE is a modified LSE by adding an extra zero argument as:
\begin{equation}
\underset{z \in \mathcal{Z}}{\mathrm{ELSE}} \, z \triangleq \underset{z \in \mathcal{Z} \cup \{ 0 \}}{\mathrm{LSE}} z = \mathrm{log}(1 + \sum\limits_{z \in \mathcal{Z}}e^z).
\label{eq:6}
\end{equation}
Since ELSE can be seen as a composite function of softplus and LSE taking their advantages simultaneously, it is adopted in various losses \cite{sohn2016improved, wang2019multi, aziere2019ensemble, kim2020proxy} with the following formulation:
\begin{equation}
\begin{split}
\mathcal{L} &= \sum_{i = 1}^{N} \left( \frac{1}{\alpha} \mathrm{log} \biggl( 1 + \sum\limits_{j \in \mathcal{P}_i} e^{\alpha (\lambda - S_{ij})} \biggr) \right.\\
& \qquad \qquad + \left. \frac{1}{\beta} \mathrm{log} \biggl( 1 + \sum\limits_{k \in \mathcal{N}_i} e^{\beta (S_{ik} - \lambda)} \biggr) \right)\\
= \sum_{i = 1}^{N} &\left( \frac{1}{\alpha} \underset{j \in \mathcal{P}_i}{\mathrm{ELSE}} \, \alpha (\lambda - S_{ij}) + \frac{1}{\beta} \underset{k \in \mathcal{N}_i}{\mathrm{ELSE}} \, \beta (S_{ik} - \lambda) \right) .
\end{split}
\label{eq:7}
\end{equation}
Then the gradients are given as:
\begin{equation}
\left\{ \begin{aligned}
\frac{\partial \mathcal{L}}{\partial S_{ij}} = & \frac{- e^{- \alpha S_{ij}}}{e^{- \alpha \lambda} + \sum\limits_{j' \in \mathcal{P}_i} e^{- \alpha S_{ij'}}}, & j \in \mathcal{P}_i\\
\frac{\partial \mathcal{L}}{\partial S_{ik}} = & \frac{e^{\beta S_{ik}}}{e^{\beta \lambda} + \sum\limits_{k' \in \mathcal{N}_i} e^{\beta S_{ik'}}}, & k \in \mathcal{N}_i\\
\end{aligned} \right. .
\label{eq:8}
\end{equation}

When we ignore terms of $\lambda$ in denominators, then the gradients turn into the softmax form.
It means that ELSE utilizes the \textit{relative-hardness} within pairs of the same set as opposed to MSP, so that the softly graded weight is assigned to each pair.
%
\section{Proxy-based framework}
\label{sec:3}
In this section, we explain a systematized methodology to expand the pair-based framework into a proxy-based framework.
\subsection{Multi-view learning and proxy assignment}
\label{ssec:3.1}
A static proxy assignment method used in \cite{movshovitz2017no, kim2020proxy} is to define proxies as much as the number of word classes with randomly initialized vectors.
As they are small enough to be stored in memory, sampling the corresponding proxies is quite simple and becomes efficient for batch construction.
However, because of independence among the vectors, it is hard to directly learn the complex relationship existing in the original embedding space.
Also, the fixed number of proxies cannot cope with unseen data classes, which is a critical issue in our field.

In contrast, we adopt multi-view learning \cite{he2017multi, jung2019additional} for the proxy assignment into our task, where proxies are generated from another network.
With a newly defined multi-view batch, $\left\{ \left( x_i, t_i, c_i \right) \right\}$, a text label $t_i$ is projected onto the embedding space by text encoder $g(\cdot; \phi)$, resulting in AGWE $g(t_i; \phi)$.
The AGWEs are discrete and exist as much as the same number of word classes, since the input text label is unique for each word class \cite{jung2019additional}.
So, we equate these generated AGWEs with proxies, and also we can still follow the static proxy assignment.
\subsection{Similarity matrix}
\label{ssec:3.2}
To establish the proxy-based framework, we have to formulate the data-to-proxy relationship.
It can be achieved by simple modification in computing the similarities between pairs.

When we substitute one element from a single-view pair $(f(x_i; \theta), f(x_j; \theta))$ with an AGWE, then there can be two possible multi-view pairs as $(f(x_j; \theta), g(t_j; \phi))$ and $(g(t_i; \phi), f(x_j; \theta))$.
In accordance, there can be defined two types of the multi-view similarity matrix as below:
\begin{equation}
\begin{aligned}
\text{Type P/N:} \quad & \mathbf{S} \triangleq \big\{ S^{\text{P/N}}_{ij} | S^{\text{P/N}}_{ij} = \mathrm{cos}(f(x_i; \theta), g(t_j; \phi)) \big\}\\
\text{Type A:} \quad & \mathbf{S} \triangleq \big\{ S^{\text{A}}_{ij} | S^{\text{A}}_{ij} = \mathrm{cos}(g(t_i; \phi), f(x_j; \theta)) \big\}\\
\end{aligned} .
\label{eq:9}
\end{equation}

The first type uses the proxies as positives or negatives (type P/N), whereas the second type uses the proxies as anchors (type A).
As shown in Eq.~(\ref{eq:9}), the resulting matrices are transposed form of each other.
It means that they can share the same matrix.
\subsection{General formulation with proxy}
\label{ssec:3.3}
We have presented constituents for establishing the proxy-based framework.
However, function $\mathcal{F}$ and multi-view similarity matrix $\mathbf{S}$ are insufficient to generalize the proxy-based framework.

We assume that the optimal function making anchor and positive closer may differ from one for anchor and negative to be farther.
It conflicts with most works that have used the symmetric formula like in Eq.~(\ref{eq:2}).
For example, in the case of MSP and ELSE introduced in Sec.~\ref{sec:2}, they show opposite characteristics of self-hardness and relative-hardness.
If we apply them together, they can create synergy while keeping their merits.

Another assumption is that proxies have different impacts when they substitute for anchor, positives, or negatives.
So, depending on which type of multi-view similarity in Eq.~(\ref{eq:9}) is adopted and whether it is applied to anchor-positive pairs or anchor-negative pairs, the role of the proxies in the training process can change.

In sum, following the above assumptions, we argue that $\mathcal{F}$ and $\mathbf{S}$ should be chosen in asymmetric combinations for the anchor-positive and anchor-negative terms.
Then, the generalized formulation for proxy-based losses is defined as:
\begin{equation}
\mathcal{L} = \sum_{i = 1}^{N} \Big( \mathcal{F}_\mathcal{P}(\mathbf{S}^\mathcal{P}, \mathcal{P}_i) + \mathcal{F}_\mathcal{N}(\mathbf{S}^\mathcal{N}, \mathcal{N}_i) \Big),
\label{eq:10}
\end{equation}
where $\mathcal{F}_\mathcal{P}(\cdot, \mathcal{P})$ and $\mathcal{F}_\mathcal{N}(\cdot, \mathcal{N})$ can be different functions that meet Eq.~(\ref{eq:3}), and $\mathbf{S}^\mathcal{P}$and $\mathbf{S}^\mathcal{N}$can be one of types in Eq.~(\ref{eq:9}).
\subsection{Asymmetric-proxy loss}
\label{ssec:3.4}
Under the proxy-based framework, a question arises of which asymmetric combination can give better performance.

From the comparative study in Sec.~\ref{ssec:4.6}, we propose to choose ELSE function of type A similarities for the anchor-positive terms and MSP function of type P/N similarities for the anchor-negative terms.
The resulting loss, which we call as \textit{Asymmetric-Proxy} loss, is given as:
\begin{equation}
\mathcal{L}^* = \sum_{i = 1}^{N} \biggl( \frac{1}{\alpha} \underset{j \in \mathcal{P}_i}{\mathrm{ELSE}} \, \alpha (\lambda - S^{\text{A}}_{ij}) + \underset{k \in \mathcal{N}_i}{\mathrm{MSP}} \, \beta(S^{\text{P/N}}_{ik} - \lambda) \biggr).
\label{eq:11}
\end{equation}
%
\section{Experiments}
\label{sec:4}
\subsection{Dataset}
\label{ssec:4.1}
We employ Wall Street Journal (WSJ) \cite{paul1992design} corpus where all utterances are segmented into word-level samples by forced alignment.
For training, we use SI-284 subset composed of 639,501 samples from 13,386 words.
For evaluation, we merge several subsets: \{dev93, dev93-5k\} into development set and \{eval92, eval92-5k, eval93, eval93-5k\} into test set.
Each has 16,839 and 18,274 samples, where the number of words is 3,289 and 3,239.
\subsection{Implementation details}
\label{ssec:4.2}
We use PyTorch \cite{paszke2019pytorch} for implementation.
Speech segments are transformed into 40 log-mel filterbank energies every 25 ms window with a 10 ms stride.
Given a text label as a character one-hot sequence, it is further transformed into 26-dimensional feature sequence by a trainable dictionary.
For the acoustic and text encoders, we use the same network architecture \cite{he2017multi, jung2019additional} that is 2-layer bi-LSTM with 512 units per direction.
And the last bidirectional outputs are concatenated to form an embedding $(d = 1024)$.
Dropout of 0.4 is applied only for the acoustic encoder.
We use the Adam \cite{kingma2015adam} optimizer with learning rate of 0.0001.
The batch size is set to 256 and two GTX 1080 Ti GPUs are used.
Models are trained for 150 epochs and we choose the model resulting the highest development set performance.
We set the hyper-parameters $\alpha = 2$, $\beta = 50$, and $\lambda = 0.5$.
All experiments are repeated 5 times and averaged for reliability.
\subsection{Evaluation tasks}
\label{ssec:4.3}
We evaluate learned AWEs and AGWEs on two word discrimination tasks: acoustic and cross-view.
Acoustic word discrimination task determines whether a pair of AWEs belongs to the same word, while cross-view word discrimination determines whether given AWE and AGWE correspond to the same word.
The decision is made by thresholding cosine similarities between embeddings.
For the evaluation metric, we use average precision (AP) that measures the area under the precision-recall curve computed by varying the threshold.
\subsection{Baselines}
\label{ssec:4.4}
We compare the results of the following baselines in our setting.
Similarities between raw speech features are measured by the dynamic time warping (DTW) algorithm.
We consider the fundamental baselines such as Contrastive, Triplet, and MV Triplet ($obj^0+obj^2$ in \cite{he2017multi}).
For Triplet and MV Triplet, we sample positives and negatives from the entire dataset.

We also compare symmetric proxy-based losses such as Proxy-NCA, Proxy-BD, and Proxy-MS which are reformulated from conventional pair-based losses \cite{goldberger2004neighbourhood}, \cite{yi2014deep}, and \cite{wang2019multi}, respectively.
We present two results for each loss according to the types of $\mathbf{S}$, which include recent proxy-based works \cite{movshovitz2017no, kim2020proxy}.
\subsection{Results}
\label{ssec:4.5}
The evaluation results are summarized in Table \ref{tab:1}.
Our proposed Asymmetric-Proxy loss outperforms all baselines on acoustic word discrimination task.
Although we observe that proxy-based methods except for one of Proxy-NCA surpass the state-of-the-art MV Triplet method in large margins, our method exceeds MV Triplet by 0.088 and the best score of baselines by 0.013 in acoustic AP.
Furthermore, our method achieves competitive performance on cross-view word discrimination task showing little difference against the best score of baselines in cross-view AP even to three decimal places.
These results demonstrate the superiority of our Asymmetric-Proxy loss.

In the case of Proxy-NCA based on LSE function, it shows relatively low performance compared to other proxy-based losses.
This implies that it is effectual to expand the triplet into the N-tuple using $\mathcal{P}$ and $\mathcal{N}$, but it is also crucial to choose functions that can utilize the margin $\lambda$.
In Eq.~(\ref{eq:8}), the margin term acts as a regularization that lowers the variance of gradients and prevents the optimization from being biased toward the hardest pair too much, while still considering their relative-hardness.
For this reason, we consider MSP and ELSE as candidates.
\begin{table}[!tp]
\caption{Word discrimination results on WSJ corpus.}
\label{tab:1}
\centering
\setlength\tabcolsep{3pt}
\begin{tabular}{c|cc|cc|c|c}
\Xhline{2\arrayrulewidth}
\multirow{2}*{\makecell{Methods}} & \multirow{2}*{\makecell{$\mathcal{F}_{\mathcal{P}}$}} & \multirow{2}*{\makecell{$\mathcal{F}_{\mathcal{N}}$}} & \multirow{2}*{\makecell{$\mathbf{S}^{\mathcal{P}}$}} & \multirow{2}*{\makecell{$\mathbf{S}^{\mathcal{N}}$}} & \multirow{2}*{\makecell{Acoustic\\AP}} & \multirow{2}*{\makecell{Cross-view\\AP}}\\
 & & & & & & \\
\Xhline{2\arrayrulewidth}
DTW & \multicolumn{2}{c|}{-} & \multicolumn{2}{c|}{-} & .099 & -\\
Contrastive & \multicolumn{2}{c|}{-} & \multicolumn{2}{c|}{-} & .479 & -\\
Triplet & \multicolumn{2}{c|}{-} & \multicolumn{2}{c|}{-} & .812 & -\\
MV Triplet & \multicolumn{2}{c|}{-} & \multicolumn{2}{c|}{-} & .833 & .910\\
\hline
\multirow{2}{*}{Proxy-NCA} & \multicolumn{2}{c|}{\multirow{2}{*}{LSE}} & \multicolumn{2}{c|}{P/N} & .869 & .927\\
 & \multicolumn{2}{c|}{} & \multicolumn{2}{c|}{A} & .815 & .894\\
\hline
\multirow{2}{*}{Proxy-BD} & \multicolumn{2}{c|}{\multirow{2}{*}{MSP}} & \multicolumn{2}{c|}{P/N} & .905 & .954\\
 & \multicolumn{2}{c|}{} & \multicolumn{2}{c|}{A} & .908 & .956\\
 \hline
\multirow{2}{*}{Proxy-MS} & \multicolumn{2}{c|}{\multirow{2}{*}{ELSE}} & \multicolumn{2}{c|}{P/N} & .908 & .963\\
 & \multicolumn{2}{c|}{} & \multicolumn{2}{c|}{A} & .881 & $\mathbf{.964}$\\
\Xhline{2\arrayrulewidth}
\rowcolor{Gray}
Ours & ELSE & MSP & A & P/N & $\mathbf{.921}$ & .963\\
\Xhline{2\arrayrulewidth}
\end{tabular}
\end{table}
\begin{table}[!tp]
\caption{Effect of asymmetry on $\mathcal{F}_{\mathcal{P}}$ and $\mathcal{F}_{\mathcal{N}}$.}
\label{tab:2}
\centering
\setlength\tabcolsep{3pt}
\begin{tabular}{cc|c|c|c}
\Xhline{2\arrayrulewidth}
\multirow{2}*{\makecell{$\mathcal{F}_{\mathcal{P}}$}} & \multirow{2}*{\makecell{$\mathcal{F}_{\mathcal{N}}$}} & \multirow{2}*{\makecell{$\mathbf{S}$}} & \multirow{2}*{\makecell{Acoustic\\AP}} & \multirow{2}*{\makecell{Cross-view\\AP}}\\
 & & & & \\
\Xhline{2\arrayrulewidth}
MSP & ELSE & \multirow{2}*{\makecell{P/N}} & .910 & .960\\
\cellcolor{Gray}{ELSE} & \cellcolor{Gray}{MSP} & & .911 & .957\\
\hline
MSP & ELSE & \multirow{2}*{A} & .874 & .959\\
\cellcolor{Gray}{ELSE} & \cellcolor{Gray}{MSP} & & .917 & .962\\
\Xhline{2\arrayrulewidth}
\end{tabular}
\end{table}
\begin{table}[!tp]
\caption{Effect of asymmetry on $\mathbf{S}^{\mathcal{P}}$and $\mathbf{S}^{\mathcal{N}}$.}
\label{tab:3}
\centering
\setlength\tabcolsep{3pt}
\begin{tabular}{c|cc|c|c}
\Xhline{2\arrayrulewidth}
\multirow{2}*{\makecell{$\mathcal{F}$}} & \multirow{2}*{\makecell{$\mathbf{S}^{\mathcal{P}}$}} & \multirow{2}*{\makecell{$\mathbf{S}^{\mathcal{N}}$}} & \multirow{2}*{\makecell{Acoustic\\AP}} & \multirow{2}*{\makecell{Cross-view\\AP}}\\
 & & & & \\
\Xhline{2\arrayrulewidth}
\multirow{2}*{\makecell{MSP}} & P/N & A & .909 & .956\\
 & \cellcolor{Gray}{A} & \cellcolor{Gray}{P/N} & .911 & .958\\
\hline
\multirow{2}*{\makecell{ELSE}} & P/N & A & .875 & .962\\
 & \cellcolor{Gray}{A} & \cellcolor{Gray}{P/N} & .915 & .962\\
\Xhline{2\arrayrulewidth}
\end{tabular}
\end{table}
\begin{table}[!tp]
\caption{Effect of asymmetry on all constituents.}
\label{tab:4}
\centering
\setlength\tabcolsep{3pt}
\begin{tabular}{cc|cc|c|c}
\Xhline{2\arrayrulewidth}
\multirow{2}*{\makecell{$\mathcal{F}_{\mathcal{P}}$}} & \multirow{2}*{\makecell{$\mathcal{F}_{\mathcal{N}}$}} & \multirow{2}*{\makecell{$\mathbf{S}^{\mathcal{P}}$}} & \multirow{2}*{\makecell{$\mathbf{S}^{\mathcal{N}}$}} & \multirow{2}*{\makecell{Acoustic\\AP}} & \multirow{2}*{\makecell{Cross-view\\AP}}\\
 & & & & & \\
\Xhline{2\arrayrulewidth}
MSP & ELSE & P/N & A & .872 & .956\\
MSP & ELSE & \cellcolor{Gray}{A} & \cellcolor{Gray}{P/N} & .906 & .958\\
\cellcolor{Gray}{ELSE} & \cellcolor{Gray}{MSP} & P/N & A & .911 & .957\\
\Xhline{2\arrayrulewidth}
\cellcolor{Gray}{ELSE} & \cellcolor{Gray}{MSP} & \cellcolor{Gray}{A} & \cellcolor{Gray}{P/N} & \cellcolor{Gray}{.921} & \cellcolor{Gray}{.963}\\
\Xhline{2\arrayrulewidth}
\end{tabular}
\end{table}
\subsection{Comparative study}
\label{ssec:4.6}
To find out the optimal asymmetric combination, we conduct a comparative study.
We investigate all possible candidates for Asymmetric-Proxy loss by alternating the constituents in Sec.~\ref{sec:3}.

In Table \ref{tab:2}, we analyze the effect of asymmetry on functions $\mathcal{F}_{\mathcal{P}}$ and $\mathcal{F}_{\mathcal{N}}$ while fixing the type of $\mathbf{S}$.
In the cases of type P/N, there is little difference in acoustic AP and cross-view AP, whereas the cases of type A show a large gap in acoustic AP.
This is because improvement with $(\mathcal{F}_{\mathcal{P}}, \mathcal{F}_{\mathcal{N}}) = (\text{ELSE}, \text{MSP})$ and degradation with $(\text{MSP}, \text{ELSE})$ occur at the same time.
The improvement to 0.917 nonetheless exceeds the baselines in Table \ref{tab:1}, so it is better to choose $(\text{ELSE}, \text{MSP})$ than other combinations including the symmetric ones.

Conversely in Table \ref{tab:3}, we fix the function $\mathcal{F}$ to analyze the effect of asymmetry on types of multi-view similarity matrices $\mathbf{S}^\mathcal{P}$and $\mathbf{S}^\mathcal{N}$.
In the cases of using MSP function, there is a slight improvement in acoustic AP and cross-view AP when $\mathbf{S}^{\mathcal{P}}$is type A and $\mathbf{S}^{\mathcal{N}}$is type P/N.
However, the cases of using ELSE function show a large gap in both tasks.
Like the results of Table \ref{tab:2}, this is because of the simultaneous improvement and degradation in performance.
Including the improvement to 0.915, $\mathbf{S}^{\mathcal{P}}$of type A and $\mathbf{S}^{\mathcal{N}}$of type P/N present consistently better aspect.

We can confirm the same trends in Table \ref{tab:4} where all constituents are considered asymmetrically.
When suggested combinations following the findings from Tables \ref{tab:2} and \ref{tab:3} are applied together, we get the best results as mentioned in Sec.~\ref{ssec:4.5}.

In the cases about the performance degradations reported in entire experiments, they show a common feature.
We presume that combinations arouse adverse effects accidentally when using $\mathcal{F}_{\mathcal{N}}$ as ELSE function of type A similarities regardless of whether their formulations are symmetric or asymmetric.
\subsection{Convergence speed}
\label{ssec:4.7}
\begin{figure}[pt]
\centering
\includegraphics[width=\columnwidth]{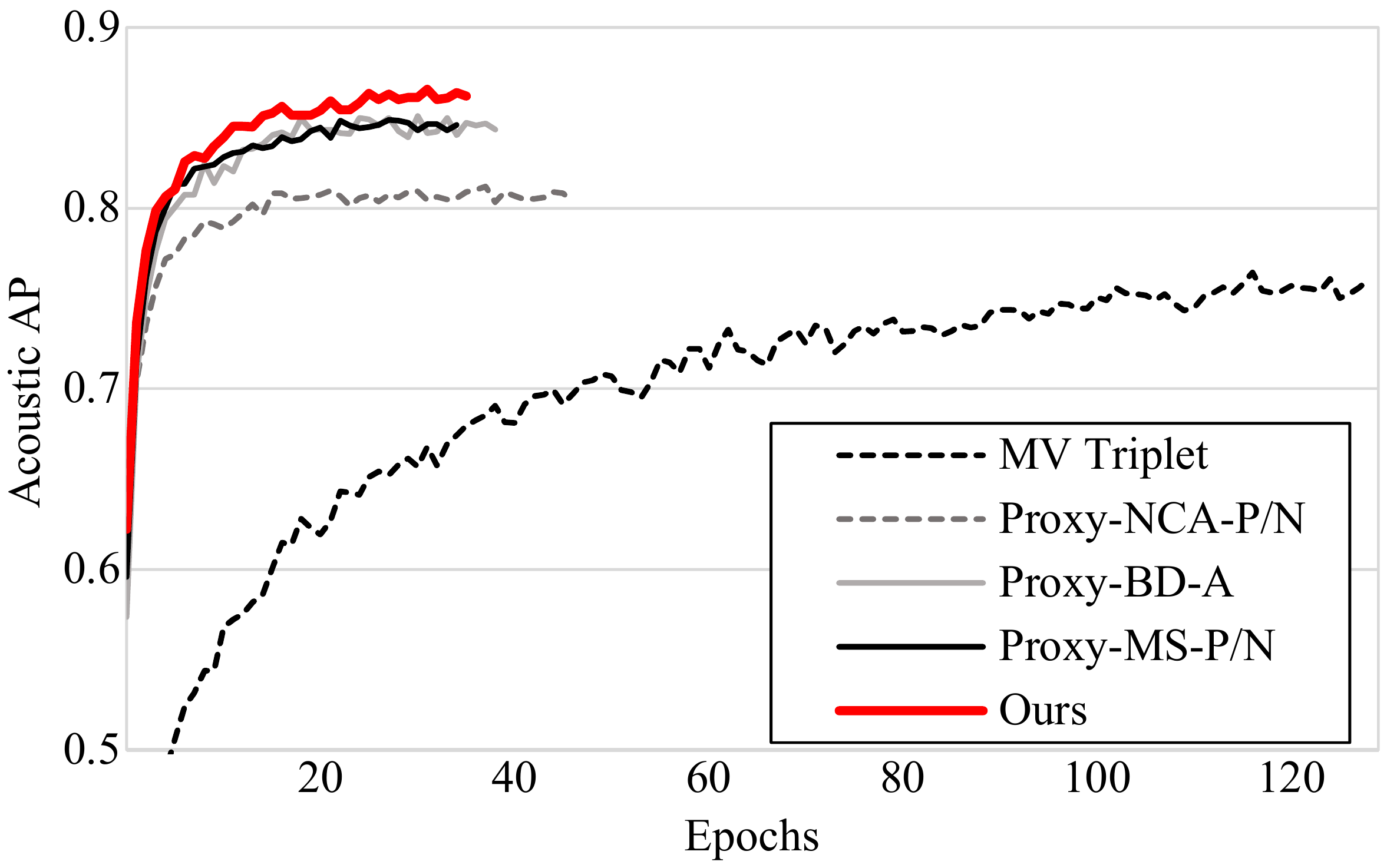}
\caption{Acoustic AP on development set along training epochs. All methods are plotted up to their convergence points.}
\label{fig:1}
\end{figure}

Since there is no additional parameters to be implemented, all methods under the proxy-based framework have the same training complexity.
Figure \ref{fig:1} compares the convergence speed of the proposed method and some baselines.
Our Asymmetric-Proxy enables to achieve the highest performance and also converges as faster as Proxy-MS with type P/N similarities, followed by Proxy-BD with type A similarities.
%
\section{Conclusion}
\label{sec:5}
Beyond expanding the multi-view learning of AWEs and AGWEs, we have established a proxy-based framework that systematizes existing proxy-based metric learning losses according to the composing functions and the types of similarity matrices.
We have further generalized the formulation based on the argument that asymmetry in combining these constituents has a meaningful effect.
Using different functions and similarity matrices for anchor-positive terms and anchor-negative terms presents a direction to improve various losses which were confined to symmetric forms.
As a result of finding the optimal combination, we proposed an Asymmetric-Proxy loss.
Our method has achieved new state-of-the-art performance on word discrimination tasks and shows consistent trends in the comparative study.
In the next work, we will investigate embedding distributions along the training progress to improve generalization performance for unseen words.
%
%
\bibliographystyle{IEEEtran}
\bibliography{mybib}

\begin{thebibliography}{10}
\providecommand{\url}[1]{#1}
\csname url@samestyle\endcsname
\providecommand{\newblock}{\relax}
\providecommand{\bibinfo}[2]{#2}
\providecommand{\BIBentrySTDinterwordspacing}{\spaceskip=0pt\relax}
\providecommand{\BIBentryALTinterwordstretchfactor}{4}
\providecommand{\BIBentryALTinterwordspacing}{\spaceskip=\fontdimen2\font plus
\BIBentryALTinterwordstretchfactor\fontdimen3\font minus
  \fontdimen4\font\relax}
\providecommand{\BIBforeignlanguage}[2]{{%
\expandafter\ifx\csname l@#1\endcsname\relax
\typeout{** WARNING: IEEEtran.bst: No hyphenation pattern has been}%
\typeout{** loaded for the language `#1'. Using the pattern for}%
\typeout{** the default language instead.}%
\else
\language=\csname l@#1\endcsname
\fi
#2}}
\providecommand{\BIBdecl}{\relax}
\BIBdecl

\bibitem{mikolov2013distributed}
T.~Mikolov, I.~Sutskever, K.~Chen, G.~S. Corrado, and J.~Dean, ``Distributed
  representations of words and phrases and their compositionality,'' in
  \emph{Advances in Neural Information Processing Systems (NeurIPS)}, 2013, pp.
  3111--3119.

\bibitem{levin2013fixed}
K.~Levin, K.~Henry, A.~Jansen, and K.~Livescu, ``Fixed-dimensional acoustic
  embeddings of variable-length segments in low-resource settings,'' in
  \emph{IEEE Workshop on Automatic Speech Recognition and Understanding
  (ASRU)}, 2013, pp. 410--415.

\bibitem{chung2016audio}
Y.-A. Chung, C.-C. Wu, C.-H. Shen, H.-Y. Lee, and L.-S. Lee, ``Audio word2vec:
  Unsupervised learning of audio segment representations using
  sequence-to-sequence autoencoder,'' in \emph{Proceedings of Annual Conference
  of the International Speech Communication Association (INTERSPEECH)}, 2016,
  pp. 765--769.

\bibitem{kamper2016deep}
H.~Kamper, W.~Wang, and K.~Livescu, ``Deep convolutional acoustic word
  embeddings using word-pair side information,'' in \emph{IEEE International
  Conference on Acoustics, Speech and Signal Processing (ICASSP)}, 2016, pp.
  4950--4954.

\bibitem{settle2016discriminative}
S.~Settle and K.~Livescu, ``Discriminative acoustic word embeddings: Recurrent
  neural network-based approaches,'' in \emph{IEEE Spoken Language Technology
  Workshop (SLT)}, 2016, pp. 503--510.

\bibitem{lim2019interlayer}
H.~Lim, Y.~Kim, J.~Goo, and H.~Kim, ``Interlayer selective attention network
  for robust personalized wake-up word detection,'' \emph{IEEE Signal
  Processing Letters}, vol.~27, pp. 126--130, 2020.

\bibitem{bengio2014word}
S.~Bengio and G.~Heigold, ``Word embeddings for speech recognition,'' in
  \emph{Proceedings of Annual Conference of the International Speech
  Communication Association (INTERSPEECH)}, 2014, pp. 1053--1057.

\bibitem{yuan2018learning}
Y.~Yuan, C.-C. Leung, L.~Xie, H.~Chen, B.~Ma, and H.~Li, ``Learning acoustic
  word embeddings with temporal context for query-by-example speech search,''
  in \emph{Proceedings of Annual Conference of the International Speech
  Communication Association (INTERSPEECH)}, 2018, pp. 97--101.

\bibitem{chen2015query}
G.~Chen, C.~Parada, and T.~N. Sainath, ``Query-by-example keyword spotting
  using long short-term memory networks,'' in \emph{IEEE International
  Conference on Acoustics, Speech and Signal Processing (ICASSP)}, 2015, pp.
  5236--5240.

\bibitem{settle2017query}
S.~Settle, K.~Levin, H.~Kamper, and K.~Livescu, ``Query-by-example search with
  discriminative neural acoustic word embeddings,'' in \emph{Proceedings of
  Annual Conference of the International Speech Communication Association
  (INTERSPEECH)}, 2017, pp. 2874--2878.

\bibitem{he2017multi}
W.~He, W.~Wang, and K.~Livescu, ``Multi-view recurrent neural acoustic word
  embeddings,'' in \emph{International Conference on Learning Representations
  (ICLR)}, 2017.

\bibitem{settle2019acoustically}
S.~Settle, K.~Audhkhasi, K.~Livescu, and M.~Picheny, ``Acoustically grounded
  word embeddings for improved acoustics-to-word speech recognition,'' in
  \emph{IEEE International Conference on Acoustics, Speech and Signal
  Processing (ICASSP)}, 2019, pp. 5641--5645.

\bibitem{hu2020multilingual}
Y.~Hu, S.~Settle, and K.~Livescu, ``Multilingual jointly trained acoustic and
  written word embeddings,'' in \emph{Proceedings of Annual Conference of the
  International Speech Communication Association (INTERSPEECH)}, 2020, pp.
  1052--1056.

\bibitem{jung2019additional}
M.~Jung, H.~Lim, J.~Goo, Y.~Jung, and H.~Kim, ``Additional shared decoder on
  {Siamese} multi-view encoders for learning acoustic word embeddings,'' in
  \emph{IEEE Workshop on Automatic Speech Recognition and Understanding
  (ASRU)}, 2019, pp. 629--636.

\bibitem{movshovitz2017no}
Y.~Movshovitz-Attias, A.~Toshev, T.~K. Leung, S.~Ioffe, and S.~Singh, ``No fuss
  distance metric learning using proxies,'' in \emph{Proceedings of the IEEE
  International Conference on Computer Vision (ICCV)}, 2017, pp. 360--368.

\bibitem{goldberger2004neighbourhood}
J.~Goldberger, G.~E. Hinton, S.~Roweis, and R.~R. Salakhutdinov,
  ``Neighbourhood components analysis,'' in \emph{Advances in Neural
  Information Processing Systems (NeurIPS)}, 2004, pp. 513--520.

\bibitem{sohn2016improved}
K.~Sohn, ``Improved deep metric learning with multi-class {N}-pair loss
  objective,'' in \emph{Advances in Neural Information Processing Systems
  (NeurIPS)}, 2016, pp. 1857--1865.

\bibitem{wang2019multi}
X.~Wang, X.~Han, W.~Huang, D.~Dong, and M.~R. Scott, ``Multi-similarity loss
  with general pair weighting for deep metric learning,'' in \emph{Proceedings
  of the IEEE/CVF Conference on Computer Vision and Pattern Recognition
  (CVPR)}, 2019, pp. 5022--5030.

\bibitem{aziere2019ensemble}
N.~Aziere and S.~Todorovic, ``Ensemble deep manifold similarity learning using
  hard proxies,'' in \emph{Proceedings of the IEEE/CVF Conference on Computer
  Vision and Pattern Recognition (CVPR)}, 2019, pp. 7299--7307.

\bibitem{kim2020proxy}
S.~Kim, D.~Kim, M.~Cho, and S.~Kwak, ``Proxy anchor loss for deep metric
  learning,'' in \emph{Proceedings of the IEEE/CVF Conference on Computer
  Vision and Pattern Recognition (CVPR)}, 2020, pp. 3238--3247.

\bibitem{dugas2000incorporating}
C.~Dugas, Y.~Bengio, F.~B{\'e}lisle, C.~Nadeau, and R.~Garcia, ``Incorporating
  second-order functional knowledge for better option pricing,'' 2000, pp.
  451--457.

\bibitem{hastie2009elements}
T.~Hastie, R.~Tibshirani, J.~H. Friedman, and J.~H. Friedman, \emph{The
  elements of statistical learning: data mining, inference, and
  prediction}.\hskip 1em plus 0.5em minus 0.4em\relax Springer, 2009, vol.~2.

\bibitem{yi2014deep}
D.~Yi, Z.~Lei, S.~Liao, and S.~Z. Li, ``Deep metric learning for person
  re-identification,'' in \emph{International Conference on Pattern
  Recognition}, 2014, pp. 34--39.

\bibitem{paul1992design}
D.~B. Paul and J.~M. Baker, ``The design for the {Wall Street Journal-based
  CSR} corpus,'' in \emph{Proceedings of the Workshop on Speech and Natural
  Language}, 1992, pp. 357--362.

\bibitem{paszke2019pytorch}
A.~Paszke, S.~Gross, F.~Massa, A.~Lerer, J.~Bradbury, G.~Chanan, T.~Killeen,
  Z.~Lin, N.~Gimelshein, L.~Antiga \emph{et~al.}, ``{PyTorch}: An imperative
  style, high-performance deep learning library,'' in \emph{Advances in Neural
  Information Processing Systems (NeurIPS)}, 2019, pp. 8024--8035.

\bibitem{kingma2015adam}
D.~P. Kingma and J.~Ba, ``Adam: A method for stochastic optimization,'' in
  \emph{International Conference on Learning Representations (ICLR)}, 2015.

\end{thebibliography}
\end{document}